\documentclass[pre,twocolumn,showpacs,aps,floats,floatfix,superscriptaddress]{revtex4}

\usepackage{graphicx}
\usepackage{subfigure}
\begin{document}

\newcommand{\avk}{\langle k \rangle}
\newcommand{\fluck}{\langle k^2 \rangle}

\title{Modeling urban street patterns}

\author{Marc Barth\'elemy} 
\affiliation{CEA-Centre d'Etudes de Bruy{\`e}res-le-Ch{\^a}tel, 
D\'epartement de Physique Th\'eorique et Appliqu\'ee BP12, 
91680 Bruy\`eres-Le-Ch\^atel, France}

\affiliation{Centre d'Analyse et Math\'ematique Sociales (CAMS, UMR 8557 
CNRS-EHESS), Ecole des Hautes Etudes en Sciences Sociales, Paris 8, France} 

\author{Alessandro Flammini} \affiliation{School of 
Informatics, Indiana University, USA}

\date{\today} 
\widetext

\begin{abstract}
  
  Urban street patterns form planar networks whose empirical
  properties cannot be accounted for by simple models such as regular
  grids or Voronoi tesselations. Striking statistical regularities
  across different cities have been recently empirically found,
  suggesting that a general and details-independent mechanism may be
  in action.  We propose a simple model based on a local optimization
  process combined with ideas previously proposed in studies of leaf
  pattern formation.  The statistical properties of this model are in
  good agreement with the observed empirical patterns. Our results
  thus suggests that in the absence of a global design strategy, the
  evolution of many different transportation networks indeed follow a
  simple universal mechanism.

\end{abstract}



\maketitle 


Transportation networks --structures that convey energy or matter from
one point to another-- appear in variety different fields, including
city streets~\cite{Cardillo,Buhl}, plant leaves~\cite{Rolland}, river
networks~\cite{Iturbe}, mammalian circulatory systems~\cite{West},
networks for commodities delivery~\cite{Gastner}, and technological
networks~\cite{Schwartz}. The recent availability of massive data sets
has opened the possibility for quantitative analysis and modeling of
these patterns and we focus here on the urban street network. Despite
the peculiar geographical, historical, social-economical mechanisms
that have shaped distinct urban areas in different ways (see for
example \cite{Levinson} and refs. therein), recent empirical studies
\cite{Batty,Makse2,Clark:1951,Crucitti,Jiang,Cardillo,Cardillo2,Lammer}
have shown that, at least at a coarse grained level, unexpected
quantitative similarities exist. The simplest description of the
street network consists of a graph whose links represent roads, and
vertices represent roads' intersections and end points. For these
graphs, links intersect essentially only at vertices and are thus
planar. Although the importance of networks in geography and urban
modeling has been recognized for a long time~\cite{Haggett},
comparably less attention has been devoted to generative models for
planar graphs in the recent literature on complex
networks~\cite{newman}. Our aim is to propose a simple model for
planar graph generation, based on plausible physical assumptions, and
which reproduces several empirical findings. In the first part of this
paper we discuss the empirical and quantitative signatures that
characterize the topology of street patterns and which suggest the
possibility to identify some general driving force steering the
formation and evolution of street patterns. In the second part, we
propose and discuss a simple and parameter-free model based on a
principle of local optimality that quantitatively reproduces the above
mentioned empirical features. The application of optimality principles
to both natural and artificial transportation networks has a long
tradition~\cite{Stevens} and in most cases requires the minimization
of a global cost function (such as the average total time for
example), in sharp contrast with the model presented here. The
rationale to invoke a local optimality principle in this context is
that every new road is built to connect a new location to the existing
road network in the most efficient way~\cite{Bejan}. The locality of
the rule is implemented both in time and space during the evolution
and formation of the street network, in order to reflect evolution
histories that greatly exceeds the time-horizon of planners. The
self-organized pattern of street emerges as a consequence of the
interplay of the geometrical disorder and the local rules of
optimality.

\begin{figure}[t!]
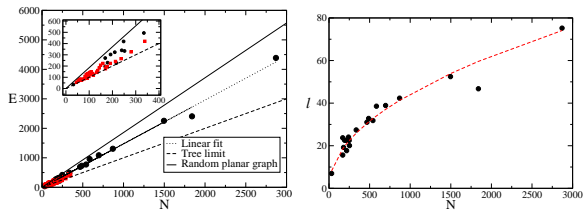

   \centering
   \includegraphics[angle=0,scale=.15]{./fig1a.eps}
   \includegraphics[angle=0,scale=.15]{./fig1b.eps}
   \caption{(a) Numbers of roads versus the number of nodes
     (ie. intersections and centers) for data from \protect\cite{Cardillo}
     (circles) and from \cite{Buhl} (squares). In the
     inset, we show a zoom for a small number of nodes. (b) Total
     length versus the number of nodes. The line is a fit which
     predicts a growth as $\sqrt{N}$ (data from \cite{Cardillo}).}
   \label{fig:k_cost}
\end{figure}

In \cite{Cardillo,Buhl} measurements for different cities in the world
are reported. Based on the data from these sources, we plot in
Fig.~\ref{fig:k_cost} the number of roads $E$ (edges) versus the
number of intersections $N$.  The plot is consistent with a linear fit
with slope $\approx 1.44$.  When individual data points are
considered, the quantity $e= E/N= \langle k \rangle / 2$ ($\langle k
\rangle$ is the average degree of a node) shows a range of values
$1.05 < e < 1.69$, in between the value $e=1$ and $e=2$ that
characterize tree-like structures and $2D$ regular lattices,
respectively. These values are however not very indicative: planarity
imposes severe constraints on the degree of a node and on its
distribution which is generally peaked around its average value. Few
exact values and bounds are available for the average degree of
classical models of planar graphs. In general it is known that $e\le
3$, while it has been recently shown~\cite{Gerke} that $e > 13/7$ for
{\it planar} Erd\"os-Renyi graphs~\cite{Gerke}.
In Fig.~\ref{fig:k_cost}b, we plot the total length $\ell$ of the
network versus N for the towns considered in~\cite{Cardillo}. Data are
well fitted by a power function of the form $\mu N^{\beta}$ with
$\mu\approx 1.51$ and $\beta\approx 0.49$).  The simplest hypothesis
consistent with the data, at this stage, is that of an homogeneous and
translational invariant structure.  Indeed, a simple scaling argument,
that could apply to a large family of planar graphs, including regular
lattices, suggests that the typical distance $\ell_1$ between
connected nodes scales as $\ell_1\sim \frac{1}{\sqrt{\rho}}$, where
$\rho=N/L^2$ is the density of vertices and $L$ the linear dimension
of the ambient space. This implies for the total length $\ell\sim
E\ell_1\sim \frac{\langle k\rangle}{2}L\sqrt{N}$. The discrepancies
between the measured $\langle k \rangle$ and $\mu$, given the error
bars, are therefore not enough to reject the hypothesis of an almost
regular lattice. However, the network of roads naturally produce a set
of non overlapping cells, encircled by the roads themselves and
covering the embedding plane, and surprisingly, the distribution of
the area $A$ of such cells measured for the city of Dresden in
Germany~\cite{Lammer} has the form $P(A)\sim A^{-\alpha}$ with
$\alpha\simeq 1.9$. This is in sharp contrast with the simple picture
of an almost regular lattice which would predict a distribution $P(A)$
very peaked around $\ell_1^2$. The authors of~\cite{Lammer} also
measured the distribution of the form factor $\phi=4A/(\pi D^2)$, (the
ratio of the area of the cell to the area of the circumscribed circle)
and found that most cells have a form factor between $0.3$ and $0.6$,
suggesting a large variety of cell shapes, in contradiction with the
assumption of an almost regular lattice. These facts thus call for a
model radically different from simple models of regular or perturbed
lattices. In the following, we describe a model where the set of
`centers' (representing new homes, businesses, etc.)  and the network
of roads that connects them grow simultaneously. New centers are
introduced every $\tau_C>1$ time steps, and for the purpose of the
present study we simply assume that the location of new points is
given exogenously and we first assume them to be randomly and
uniformly located over a square of given size. Finite segments (of
fixed and small length) of roads are simultaneously added to the
existing network every $\tau_R = 1<\tau_C$ in order to account for the
limited time horizon of planners. The algorithm that drives the
construction of new portions of roads is based on a local optimality
principle and aims at connecting to the network the still unconnected
centers using as little as possible road-length.

\begin{figure}[!t]
   \centering
   \includegraphics[angle=0,scale=.30]{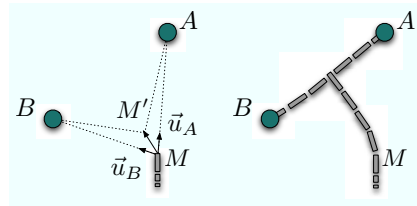}
   \caption{M is the closest network's point to both center $A$ and $B$. The
     road will grow to point M' in order to maximally reduce the cumulative 
     distance $\Delta$ of A and B from the network.}
   \label{fig:delta}
\end{figure}

In order explain the algorithm, we illustrate it on the simple example
of Fig.~\ref{fig:delta}. We assume that at a given stage of the
evolution, two centers $A$ and $B$ still need to be connected to the
network.
At any time step, each center can trigger the construction of a single
new portion of road of fixed (small) length. In order to maximally reduce 
their distance to the network, both $A$ and $B$ would select the closest points 
$M_1$ and $M_2$ in the network as initial points of the new portions of roads to be built. 
If $M_1$ and $M_2$ are distinct, segments of roads are added
along the straight lines $M_1A$ and $M_2B$. If $M_1=M_2=M$, it is not
economically reasonable to build two different segments of roads and
in this case only one single portion $MM'$ of road is allowed. Our
main assumption is that the best choice is to build it in order to
maximize the reduction of the cumulative distance from the network
($M$) to $A$ and $B$
\begin{equation}
\Delta=[d(M,A)+d(M,B)]-[d(M',A)+d(M',B)]
\end{equation}
The maximization of $\Delta$ is done under the constraint
$|MM'|= \mathrm{const.}\ll 1$ and a simple calculation leads to
\begin{equation}
\overrightarrow{MM'}\propto \vec{u}_A+\vec{u}_B
\label{vec_rule}
\end{equation}
where $\vec{u}_A$ and $\vec{u}_B$ are the unit vectors from M in the
direction of A and B respectively. The rule (\ref{vec_rule}) can
easily be extended to the situation where more than two centers want
to connect to the same point $M$. Already in this simple setting non
trivial geometrical features appear. In the example of
Fig.~\ref{fig:delta} the road from $M$ will develop a bended shape
until it reaches the line $AB$ and intersects it perpendicularly as it
is commonly observed in most urban settings. At the intersection
point, a singularity occurs with $\vec{u}_A+\vec{u}_B \approx 0$ and
one is then forced to grow two independent roads from the intersection
to A and B. The above procedure is iterated until all centers are
connected. Interestingly, although the minimum expenditure principle
was not used, the rule Eq.~(\ref{vec_rule}) was proposed by Runions
{\it et al.}~\cite{Rolland} in a study about leaf venation patterns
and we can follow their implementation. In particular, the growth
scheme described so far leads to tree-like structures and we implement
ideas proposed in~\cite{Rolland} in order to create networks with
loops. Indeed, even if tree-like structures are on one side
economical, on the other hand, they are hardly efficient (for example
the path length along a minimum spanning tree scales as a power $5/4$
of the Euclidean distance between the end-points~\cite{Duco}) and
better accessibility is granted if loops are
present. Following~\cite{Rolland}, we assume that a new center can
trigger the construction of more than one portion of road per time
step.  An unconnected center $s$ `stimulates' the addition of new
portion of roads from any vertex $v$ of the road network (vertices
correspond to any end points of the previously introduced road
segments) that is in its relative neighborhood~\cite{Toussaint}. A
node $v$ belongs to the relative neighborhood of $s$ if for any node
$u$ (center or vertex of the road network) the inequality
$d(v,s)<\max(d(s,u),d(u,v))$ holds~\cite{rel_nei}, which captures the
loosely defined requirement that $v$ belongs to the relative
neighborhood of $s$ if the region between $s$ and $v$ is
empty. Centers can therefore be reached by more than one road, leading
to the formation of loops. When more than one center stimulates the
same point the prescription of Eq.~(\ref{vec_rule}) is applied and the
evolution ends when the list of stimulated points is exhausted.
\begin{figure}[t!]
  \centering
  \includegraphics[angle=0,scale=.30]{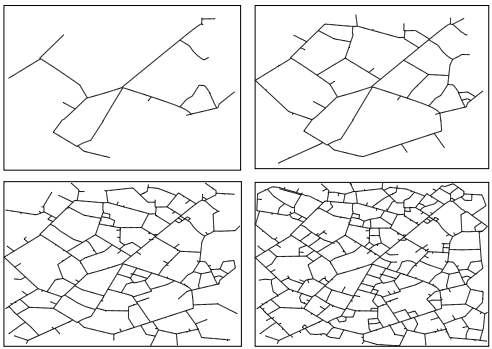}
  \caption{Snapshots of the network at different times of its
    evolution: for (a) $t=100$, (b) $t=500$, (c) $t=2000$, (d)
    $t=4000$ (the growth rate is here $\eta=0.1$). At short times, we
    have almost a tree structure and loops appear for larger density
    values obtained at larger times (the number of loops then
    increases linearly with time). 
}
   \label{fig:time}
\end{figure}

\begin{figure}[h!]
\centering
   \includegraphics[angle=0,scale=.30]{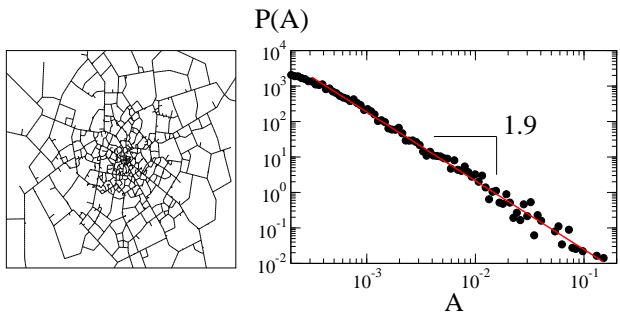}
   \caption{(a) Network obtained for an exponential distribution of
     centers ($1000$ centers and $r_c=0.1$). (b) In this case, the
     area distribution is a power law (obtained for $5000$ centers and
     $100$ configurations). The solid line is a power law fit with an
     exponent $\approx 1.9$ (for this size and with a smaller number
     of configurations, we observe fluctuations of this exponent of the
     order of $10\%$).}
   \label{fig:area}
\end{figure}

The final road network is achieved starting from a small set of $n_0$
centers connected by roads and iterating the following two steps: (i)
at every time multiple of $\tau_C$ we add $n$ new centers whose
locations $r$ are chosen randomly according to a given distribution
$P(r)$; (ii) the road network grows according to the algorithm
described above. When a center is reached by all the roads it
activated, it becomes `inactive' and cannot stimulate the growth of a
road any longer. We repeat (i-ii) until the total number of centers
reaches a desired value.  Although it is clear that the focus of the
present paper is on the road network growth, it is important to stress
that our model relies on a number of simplifying assumptions, the most
relevant of which is the fact that the centers are independently
located one from the other and from the structure of the road network.
In fact, strong evidences~\cite{Makse2,Jensen:2006} suggest that this
is not the case, and integrating the correlation centers-network is
the next most important step~\cite{inprep}. Despite this limitation,
the model produces realistic results, in good agreement with empirical
data (discussed below) which demonstrates that even in the absence of a well
defined blueprint, non-trivial global properties emerge. In
Fig.~\ref{fig:time} example of patterns obtained for a spatially
uniform distribution of new centers are shown for different times.
\begin{figure}[t!]
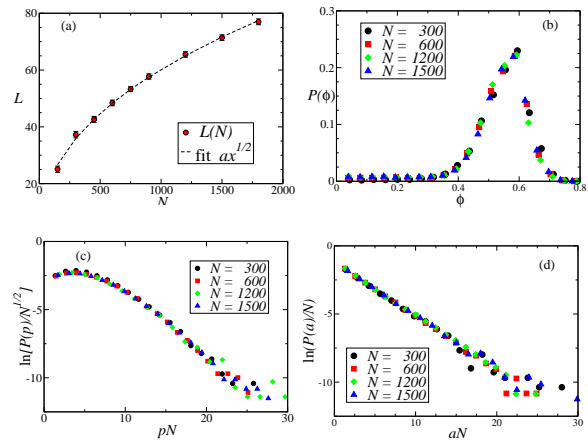

   \begin{center}
\subfigure{
   \includegraphics[angle=0,scale=.15]{./fig4a.eps}
   \includegraphics[angle=0,scale=.15]{./fig4b.eps}
}
\vspace{0.1cm}
\subfigure{
   \includegraphics[angle=0,scale=.15]{./fig4c.eps}
   \includegraphics[angle=0,scale=.15]{./fig4d.eps}
}
   \end{center}
   \caption{Simulation results (averaged over $1000$
     configurations). (a) Total length of roads versus the number of
     nodes. The dotted line is a square root fit. (b) Structure factor
     distribution showing a good agreement with the empirical results
     of~\protect\cite{Lammer}. (c-d) Rescaled distributions of the
     perimeter (c) and of the areas (d) of the cells displaying an
     exponential behavior.}
   \label{fig:l_phi}
\end{figure} 
As time progresses, density increases, and the typical length from a
center to the existing road network shortens and scales as $\ell_1\sim
1/\sqrt{\rho}\sim 1/\sqrt{t}$ as observed in the simulations. Beyond
visual similarities with real cities, the ratio $e=E/N$ has initially
a value around $1$ (corresponding to a tree-like network) and
increases very fast with $N$ reaching a value around $e\approx 1.3$
which is not far from the empirical finding (here and in the
following, we checked that the results were robusts for different
values of the growth rate $n/\tau_C$). Consistently, in
Fig~\ref{fig:l_phi}a we observe a relationship between the total
length and N that is well approximated by a function of the form
$a\sqrt{N}$ with $a\approx 1.90$, again in reasonable agreement with
the empirical data. Panels b, c, and d of Fig.~\ref{fig:l_phi} show
the collapse for different values of $N$ of the distributions of
$\phi$, $A$ and the perimeter $p$ of the cells, respectively. The
excellent collapses show that the structures obtained are consistent
with the hypothesis of homogeneity and translational invariance
formulated above. We also note that the distribution of the $\phi$
factor is peaked around $0.6$ and essentially supported in the
interval $0.4 < \phi < 0.7$, in very good agreement with facts
reported earlier~\cite{Lammer}. A simple spatial uniform disorder and
a plausible mechanism that connects the centers to the network can
thus explain the non-trivial form of the $\phi$ factor distribution,
but predicts an exponential behavior for the area distribution
(Fig.~\ref{fig:l_phi}d), in disagreement with empirical
observations~\cite{Lammer}. In real cases however, the density of
centers is not uniform.  We therefore relax this assumption and
assume, as supported by a previous empirical study~\cite{Clark:1951},
that the centers' distribution follows the population density and
decreases as $P(r)=\exp (-|r|/r_c)$ where $r$ is the distance from the
central business district and $1/r_c$ the population density
gradient. Although most quantities (such as $\langle k \rangle$ and
$\ell$) are not sensitive to the centers' distribution, the impact on
the area distribution is drastic. Indeed, as shown in
Fig.~\ref{fig:area}, we observe a power law decay with an exponent
equal to $1.9\pm0.05$ in remarkable agreement with the empirical
result of \cite{Lammer} for the road network of the city of Dresden.
This agreement confirms that the simple local optimization is a good
candidate for the main process driving the evolution of city street
patterns but also shows that the center spatial distribution $P(r)$ is
crucial.


More than $50\%$ of the world population lives today in cities and
this figure is bound to increase \cite{UN}. This migration effect has
dictated a fast and short-term planned urban growth which needs to be
understood and modeled in terms of socio-geographical contingencies
and of the general forces that drive the development of
cities. Previous studies about urban morphology have mostly tried to
identify specific mechanisms that have shaped distinct urban areas in
different ways. Here we studied a simple model based on the assumption
that road networks develop trying to grant in an efficient and at the
same time economic way connections to a set of `centers'. The model
accounts quantitatively for a list of descriptors that characterize
the topology of street patterns, and in a more qualitative way for the
tendency to have bended roads-even in absence of geographical
obstacles-and perpendicular intersections. Interestingly, the
optimality principle applied here turns out to be general and was
implicitly at the basis of a model previously
investigated~\cite{Rolland} to explain the formation of veins'
patterns in leaves, pointing to an unexpected generality of the
principle in the formation of transportation systems. This model is
simple enough to allow many interesting generalizations. In
particular, our results suggest that the local optimality principle is
a key ingredient for a more general model describing the co-evolution
of the center distribution and the network~\cite{inprep}.





\begin{thebibliography}{99}
\bibitem{rel_nei}
The notion of relative neighborhood can be equivalently stated 
as follows: a  node $v$ belongs to the relative neighborhood of $s$ if 
for any node $u$ such that $d(u,s) < d(u,v)$, it also holds that 
$d(v,s)<d(v,u)$.
\end{thebibliography}
\end{document}